\documentclass[a4paper]{jpconf}
\usepackage{graphicx}
\usepackage{subfigure}
\begin{document}
\title{Neutrino cross section measurements @ SciBooNE}

\author{C. Mariani}

\address{Columbia University, New York, NY 10027, USA}

\ead{mariani@nevis.columbia.edu}

\begin{abstract}
We report measurements of cross sections of neutrinos of 0.7~GeV average energy scattering off 
a carbon target cross sections with by the SciBooNE experiment at Fermilab. These measurements are
important inputs for current and future accelerator-based neutrino oscillation experiments in the
interpretation of neutrino oscillation signals.
\end{abstract}

\section{Introduction}\label{sec:introduction}

The measurement of neutrino mixing angle $\theta_{13}$ is one of the most important goals
in current neutrino experiments. For the current and next generation of long baseline
neutrino oscillation experiments, T2K, NOvA and LBNE, the precise measurement of
neutrino-nucleus cross sections in the few GeV energy range is an essential ingredient
in the interpretation of neutrino oscillation signals. 

\section{SCIBOONE EXPERIMENT}\label{sec:sciboone_exp}

The SciBooNE experiment is designed to measure neutrino cross sections on carbon
near one GeV region. The experiment collected data in 2007 and 2008 with neutrino ($0.99\times 10^{20}$ POT) and
antineutrino ($1.53\times10^{20}$ POT) beams in the FNAL Booster Neutrino Beam line (BNB). The BNB uses a
primary proton beam with kinetic energy 8~GeV and a beryllium target placed inside an
aluminum horn as a neutrino production target. 
The SciBooNE detector is located 100~m downstream from the neutrino production target. The neutrino flux in the SciBooNE
detector is dominated by muon neutrinos. The flux-averaged mean neutrino energy is
0.7~GeV in neutrino running mode and 0.6~GeV in antineutrino running mode
\par The SciBooNE detector consists of three detector components; SciBar, Electromagnetic
Calorimeter (EC) and Muon Range Detector (MRD). SciBar is a fully active and
fine grained scintillator detector that consists of 14,336 bars arranged in vertical and
horizontal planes. SciBar is capable of detecting all charged particles and performing
dE/dx-based particle identification. The EC is located downstream of SciBar. The detector
is a ÒspaghettiÓ calorimeter with thickness of 11$X_0$ and is used to measure $\pi^0$
and the intrinsic $\nu_e$ component of the neutrino beam. The MRD is located downstream
of the EC in order to measure the momentum of muons up to 1.2~GeV/c using their range.
It consists of 2-inch thick iron plates sandwiched between layers of plastic scintillator
planes.

%CC

\section{Charged Current Neutrino Interactions}

\begin{figure}[htbp!]
\begin{center}
\includegraphics[width = \columnwidth]{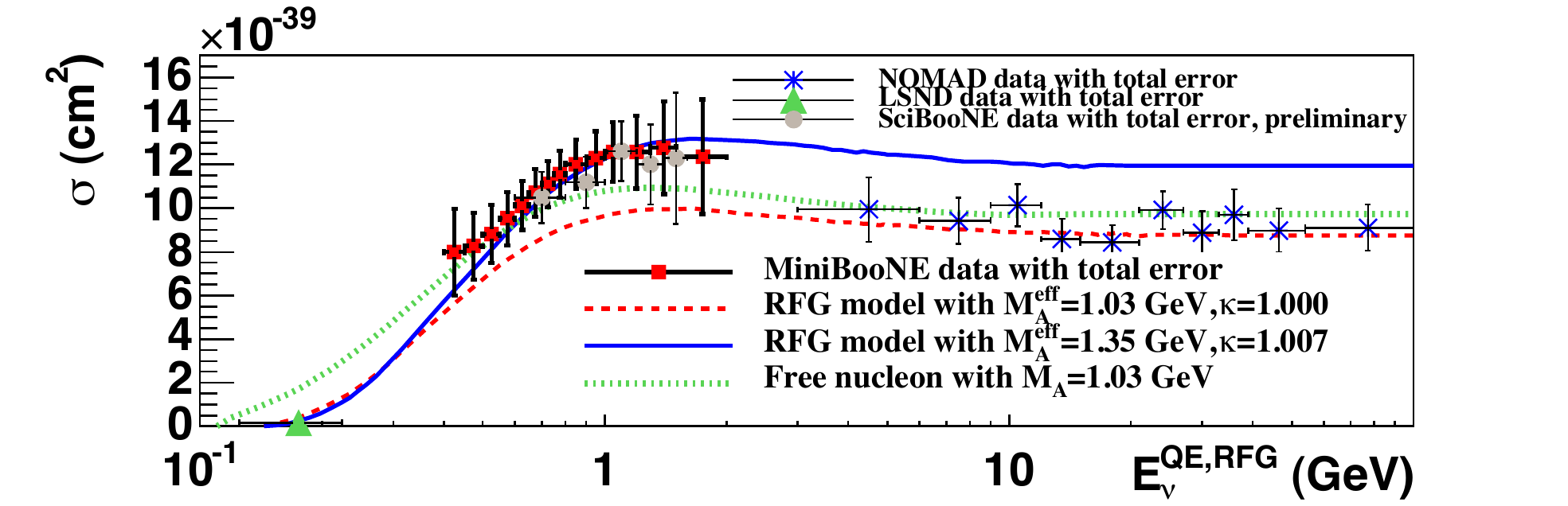}
\caption{CCQE cross-section versus neutrino
energy for the SciBooNE, MiniBooNE~\cite{AguilarArevalo:2010zc} and NOMAD~\cite{Lyubushkin:2008pe} experiments.}
\label{fig:CCQE_enu}
\end{center}
\end{figure}

The SciBooNE collaboration reported a charged current quasi-elastic (CCQE)
cross section measurement in \cite{AlcarazAunion:2009ku}. 
We developed two distinct CCQE data sets, one with tracks
matched between SciBar and the MRD and the other with tracks contained
within SciBar. To simulate neutrino-nuclear scattering, SciBooNE uses
the NEUT~\cite{Mitsuka:2008zz} generator Monte Carlo (MC) simulation. 
In the SciBar-contained sample, the muons from charged-current
neutrino candidates are tagged with particle identification based on
energy deposition along the track and by searching for their decay
electrons using SciBar's multi-hit TDCs. Events are
further classified based on the number and type of tracks coming from
the neutrino vertex. Removing the MRD track-matching cut allows the
reconstruction of backwards-going tracks, thus expanding the $Q^2$,
range open to the analysis. The neutrino vertex is defined using the
timing of hits within the muon track, and the location of the tagged
decay electron. Figure~\ref{fig:CCQE_enu} shows the SciBooNE's CCQE cross section measurement and 
compares it with MiniBooNE and NOMAD CCQE measurements.  We note that the SciBooNE and MiniBooNE results are
obtained directly from the measured event yields and
proton-on-target-normalized neutrino flux predictions, not by
extracting via a cross-section ratio with a different (assumed)
cross-section.  These are the world's first such POT-normalized
neutrino CCQE cross-section measurements.

In addition to the CCQE cross section the SciBooNE collaboration has reported in \cite{Nakajima:2010fp} a measurement of inclusive charged current interactions of muon neutrinos on carbon with an average energy of 0.7~GeV. Measuring $p_\mu$ and $\theta_\mu$, we have extracted the CC interaction rates and cross sections using flux and neutrino interaction simulations. We extract the $\nu_\mu$ CC interaction rates by fitting muon kinematics, with precision of 6-15\% for the energy dependent and  3\% for the energy integrated analyses. We confirmed that the distributions 
after fitting well reproduce the predicted distributions with both NEUT and NUANCE\cite{Casper:2002sd}
based simulations. We have also extracted CC inclusive interaction cross sections from the
observed rates, with a precision of 10-30\% for the energy dependent and 8\% for the energy integrated analyses. This is the first measurement of the CC inclusive cross section on carbon around 1~GeV.

%if there is some space maybe add the two figures from Nakaji paper using subfigure ccinclusive and ccratemeasurement

%\begin{figure}[htbp!]
%\begin{center}
%\subfigure[CC inclusive interaction cross section.]{\includegraphics[width = 0.3\columnwidth]{ccinclusive.pdf}}
%\qquad
%\subfigure[CC interaction rate normalized to the
%NEUT and NUANCE predictions.]{\includegraphics[width = 0.3\columnwidth]{ccratemeasurement.pdf}}
%\caption{Figures are taken from \cite{Nakajima:2010fp}}
%\label{fig:CCQE_enu}
%\end{center}
%\end{figure}

\section{Neutrino charged current - pion production}

\subsection{Charged Current coherent production in neutrino and antineutrino mode}

\begin{figure}[htbp!]
\begin{center}
\includegraphics[width = \columnwidth]{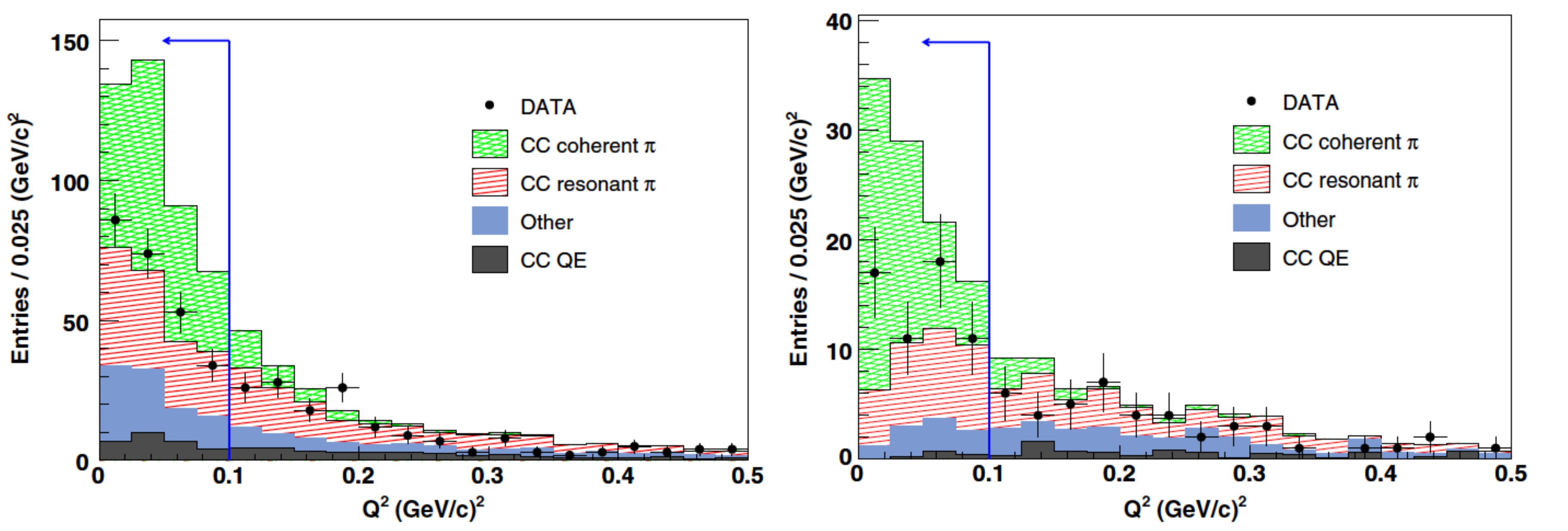}
\caption{Reconstructed $Q^2$ distributions for two distinct subsamples; mean neutrino energy of 1.1~GeV (right) and 2.3~GeV (left). Figures are from \cite{Hiraide:2008eu}.}
\label{fig:pionQ2}
\end{center}
\end{figure}

The experimental signature of charged current coherent pion production is: 1) the
existence of two and only two tracks originating from a common vertex, both consistent
with minimum ionizing particles, a muon and a charged pion, 2) no nucleon recoil in
the final state, 3) small momentum transfer. We searched for charged current coherent
pion production with two distinct data samples; averaged energy 1.1~GeV and 2.3~GeV.
Figure~\ref{fig:pionQ2} shows the $Q^2$ distribution in the coherent pion sample for the neutrino energy of 1.1~GeV (left plot) and 2.3~GeV (right plot). There is a data deficit in the low $Q^2$ region. No evidence for coherent pion production was observed. 
We evaluated the ratio of the charged current coherent pion production cross section to
the charged current inclusive cross section to be: $(0.16\pm0.17(stat)^{+0.30}_{-0.27}(sys))\times10^{-2}$ at 1.1~GeV, and $(0.68\pm0.32(stat)^{+0.39}_{-0.25}(sys))\times10^{-2}$ at 2.2~GeV~\cite{Hiraide:2008eu}. We confirmed the K2K~\cite{Hasegawa:2005td} results and we set 90\% confidence level upper limits on the cross section ratio of charged current coherent pion production to the total charged current cross section at $0.67\times10^{-2}$ at a mean neutrino energy of 1.1~GeV and $1.36\times10^{-2}$ at a mean neutrino energy of 2.2~GeV. 

SciBooNE has also performed a search for charged current coherent pion production in anti-neutrino
mode running following the analysis performed on neutrino data. The only difference from the neutrino mode analysis is in the treatment of the neutrino
(wrong-sign) contribution since SciBooNEÕs $\bar{\nu}$ beam contains a relatively
large fraction of neutrinos. The preliminary results of charged current coherent pion production cross section ratio in SciBooNEÕs anti-neutrino mode running data is obtained to be  $\frac{\sigma(\bar{\nu}CC coh-\pi)+r\cdot \sigma(\nu CC coh-\pi)}{\sigma(\bar{\nu} CC) + r \cdot \sigma(\nu CC)} = (1.131 \pm 0.34(stat)^{+0.31}_{-0.36}(sys)) \times 10^{-2}$ at the mean anti-neutrino energy of 1.0~GeV. $r=\frac{\int \Phi(\nu) dE}{\int \Phi(\bar{\nu}) dE } = 0.19$ is a ratio of anti-neutrino and neutrino fluxes in the Booster Neutrino Beam based on SciBooNEÕs flux prediction~\cite{Nakajima:2010fp}.

\subsection{Neutral pion production}

The experimental signature of charged current neutral pion production is constituted
by two electromagnetic cascades initiated by the conversion of the $\pi^0$ decay gamma
rays, with an additional muon in the final state from charged current processes. The
obtained preliminary results of charged current $\pi^0$ cross section is $5.6\pm1.9(stat)\times 10^{-40} cm^2$ per nucleon at the mean neutrino energy of 0.9~GeV. The systematic error is currently under evaluation. The measured cross section result agrees with the NEUT prediction within statistical uncertainty. 

%NC
\section{Neutral Current neutrino interactions}

The characteristic topology of neutral current $\pi^0$ production events is two gamma rays
from $\pi^0$ decay in the final state. We reconstruct gamma rays converting in SciBar and
select events with two reconstructed gamma rays and no muons. A neutral current $\pi^0$
interaction here is defined as neutral current neutrino interaction in which at least one $\pi^0$
is emitted in the final state from the target nucleus. We obtained $(7.7\pm0.5(stat)\pm0.5(sys))\times10^{-2}$ as the ratio of the neutral
current neutral pion production to total charged current cross section as in \cite{Kurimoto:2009wq}. We measured the
ratio of cross sections in order to minimize systematic uncertainty due to the neutrino
flux prediction. The result agrees with the Rein-
Sehgal model implemented in our neutrino interaction simulation program with nuclear
effects. The spectrum shape of the $\pi^0$ momentum and angle also agree with the model.
\par The SciBooNE collaboration performed a search for neutral current coherent pion production using
the neutral current $\pi^0$ sample described before. In neutral
current coherent pion production, there is no recoil nucleon in the final state since the $\pi^0$
is produced by the neutrino interacting with the whole nucleus. To separate the neutral current
coherent $\pi^0$ events from the neutral current resonant $\pi^0$ events, recoil protons in the final
state are used. We measured the ratio of the neutral current coherent $\pi^0$ production to total charged current cross sections to be $(1.16\pm0.24)\times 10^{-2}$. The error includes both the statistical and systematical errors. The ratio of charged current coherent $\pi^+$ to neutral current coherent $\pi^0$ production is calculated to be $0.14^{+0.30}_{-0.28}$, using our published charged current coherent pion measurement.

\section{$K^+$ Production at BNB}

SciBooNE reported a $K^{+}$ production cross section and rate measurements in interactions of protons on the MiniBooNE beryllium target in the BNB in \cite{Cheng:2011wq}: this was done using high energy muons produced by neutrinos interacting in the SciBar detector. Using observed neutrino and antineutrino events in SciBooNE, we measure $\frac{d^2\sigma}{dpd\Omega}~= ~(5.34\pm0.76)~mb/(GeV/c \times sr) $ for $p + Be\rightarrow K^+ + X$ at mean $K^{+}$ energy of 3.9~GeV and angle (with respect to the proton beam direction) of 3.7~degrees, corresponding to the selected $K^+$ sample.

\section{ACKNOWLEDGMENTS}
We acknowledge the support of Fermilab, MEXT, JSPS (Japan), the INFN (Italy), the Ministry of Science
and Innovation and CSIC (Spain), the STFC (UK), and the DOE and NSF (USA).

%\vspace{2 cm}
\section*{References}
\bibliographystyle{unsrt}

\bibliography{references}

\end{document}